\documentclass[copyright,creativecommons,noncommercial]{eptcs}
\providecommand{\event}{PLACES 2014} % Name of the event you are submitting to
\usepackage{breakurl}             % Not needed if you use pdflatex only.

\usepackage{amsmath}
\usepackage{amssymb}
\usepackage[mathscr]{eucal}

\usepackage{xcolor}
\usepackage{stmaryrd}
\usepackage{graphicx}
\usepackage{url}
\usepackage{mathpartir} % inference rules package
\usepackage{framed}
\usepackage{cmll}
\usepackage{mathtools}
\usepackage{accents}
\usepackage{enumerate}
\usepackage{paralist}
\usepackage{eurosym}
\def\eg{{\em e.g.}}

\usepackage{prooftree}

\makeatletter

\newskip \point

\def \premisespacing{\quad}

\def \RuleFurtherPremise[#1]#2#3#4{\RuleMultiPremise[#1]{#2\premisespacing #3}{#4}}

\def \RuleMultiPremise[#1]#2#3{\@ifnextchar\bgroup {\RuleFurtherPremise[#1]{#2}{#3}}{\prooftree #2\justifies#3 \using{#1}\endprooftree}}

\def \RuleWithName[#1]#2{\@ifnextchar\bgroup {\RuleMultiPremise[#1]{#2}}{\prooftree \justifies #2 \using{#1} \endprooftree}}

\def \RuleWithInfo[#1]{\@ifnextchar[{\RuleWithNameAndCondition[#1]}{\RuleWithName[(#1)]}}

\def \RuleWithNameAndCondition[#1][#2]{\RuleWithName[(#1)^{#2}]}

\def \Inf{\proofrulebaseline=2ex \abovedisplayskip12\point\belowdisplayskip12\point \abovedisplayshortskip8\point\belowdisplayshortskip8\point \@ifnextchar[{\RuleWithInfo}{\RuleWithName[ ]}}

\makeatother

\usepackage{tikz}
\usetikzlibrary{calc}
\usetikzlibrary{arrows}
\usetikzlibrary{positioning}
\usetikzlibrary{fadings}
\usetikzlibrary{chains}
\usetikzlibrary{fit}
\newcommand{\smm}[1]{{\ensuremath{#1}}}

\newcommand{\denv}[1]{\ensuremath{[{#1}]}}
\newcommand{\forget}[1]{}
\newcommand{\NI}{\noindent}
\newif\ifattn\attntrue
\newenvironment{th:figure}[1][tbhp]%
{\begin{figure}[#1]}{\end{figure}}%
\newenvironment{th:figurebody}%
{\begin{framed}\vspace{-2ex}}{\vspace{-2ex}\end{framed}}

\newcommand{\rul}[1]{\mbox{\small\sf (#1)}}

\newcommand{\ie}{i.e.}
\newcommand{\set}[1]{\ensuremath{\left\{#1\right\}}}

\newcommand{\ovl}[1]{\ensuremath{\overline{#1}}}

\newcommand{\defeq}{\ensuremath{\,\doteq\,}}

\newcommand{\nil}{\ensuremath{\mathbf{0}}}

\newcommand{\subs}[2]{\ensuremath{ \{ \raisebox{2.6pt}{{\small $ #1 $}} \!\, / \!\, \raisebox{-1.0pt}{{\small $#2$}} \} }}

\newcommand{\fn}[1]{\ensuremath{ \mathsf{fn}(#1) }}
\newcommand{\acn}[1]{\ensuremath{ \mathsf{an}(#1) }}

\newcommand{\an}[1]{\acn{#1}}

\newcommand{\pn}[1]{\ensuremath{ \mathsf{pn}(#1) }}
\def\Rq{\:\R\:}

\def\osred{\longrightarrow}             % one-step reduction relation
\def\R{\osred}

\newcommand{\kw}[1]{\ensuremath{\mathsf{#1}}}
\newcommand{\tp}{\ensuremath{\colon}}
\newcommand{\tps}{\ensuremath{{:}}}

\newcommand{\dual}[1]{\ensuremath{\lneg #1}}

\newcommand{\sesst}{S}   % The session type
\newcommand{\Ga}{\ensuremath{\Gamma}}

\newcommand{\De}{\ensuremath{\Delta}}

\newcommand{\out}[1]{\ensuremath{\langle#1\rangle}}

\newcommand{\myeqn}[2]%
{\ensuremath{#2} \quad \llabel{#1}}

\newenvironment{sr-case}[3]%
{\locallabelreset\NI \mbox{{\bf\em Case}} \rul{#1} \ensuremath{\qquad #2 \qquad #3} \vspace{6pt}\newline%
\NI %\hspace*{0.03\textwidth} \begin{minipage}{0.97\textwidth}  
}
{%\end{minipage}
}

\newcommand{\var}[1]{\textit{#1}}

\newcommand{\ftv}[1]{\ensuremath{ \mathsf{ftv}(#1) }}
\newcommand{\newses}[1]{\ensuremath{ \kw{session}: ( \sesst ) }}

\newcommand{\prl}{\ensuremath{\; | \;}}
\newcommand{\prls}{\ensuremath{\, | \,}}
\newcommand{\bprl}{\ensuremath{\; \big| \;}}

\newcommand{\new}{\ensuremath{\boldsymbol \nu}}
\newcommand{\newc}[1]{\ensuremath{(\new #1)}}

\newcommand{\munit}{\ensuremath{\mathbf{1}}}

\newcommand{\sat}{\mathop{\scalebox{0.6}{\raisebox{0.0ex}{\ensuremath{\!\textbf{@}\!}}}}}%

\newcommand{\hast}{\ensuremath{\:\triangleright\:}}

\newcommand{\ctx}[1]{\ensuremath{\mathsf{C}[\,#1\,]}}

\newcommand{\qt}[1]{``{#1}''}

\newcommand{\PNS}{proof nets}

\newcommand{\LL}{Linear Logic}

\newcommand{\aconv}{\ensuremath{\alpha}-conversion}

\newcommand{\tsf}[1]{\textsf{#1}}

\newcommand{\choiceI}[3]{\ensuremath{ \ovl #1 \triangleleft {\bfs #2} (#3) }}

\newcommand{\branchIT}[6]{\ensuremath{#1 \triangleright \set{ {\bfs #2}( #3_{#2} \tps #4_{#2} ).#5_{#2} }_{#2 \in #6}}}
\newcommand{\branch}[2]{\ensuremath{#1 \triangleright \set{#2}}}

\newcommand{\orBra}{\ensuremath{ \talloblong }}
\newcommand{\methI}[3]{\ensuremath{ {\bfs #1} (#2).#3 }}

\newcommand{\wni}{\ensuremath{\wn {\!\!\:}_\star}}
\newcommand{\oci}{\ensuremath{\oc_\star}}

\newcommand{\wnm}{\ensuremath{\wn {\!\!\:}_{\md}}}

\newcommand{\ocm}{\ensuremath{\oc_{\md}}}

\newcommand{\ocmi}[1]{\ensuremath{\oc_{\md_{#1}}}}

\newcommand{\mde}{\ensuremath{{\boldsymbol \varepsilon}}}
\newcommand{\mdi}{\ensuremath{\star}}
\newcommand{\md}{\ensuremath{\mathfrak{m}}}

\newcommand{\bfs}[1]{\ensuremath{\boldsymbol #1}}

\newcommand{\lneg}{\ensuremath{{\thicksim}}}

\renewcommand{\bot}{{\mathop{\tikz[baseline=(neg.base)]{ %\clip (-4pt,-3.3pt) rectangle (4pt,4pt) ; 
  \pgfsetbaseline{-0.1ex}
  \node[inner sep=0pt, outer sep=0pt] at (0pt,0pt) (neg) {\lneg} ; 
 \path[draw,line width=0.22ex,inner sep=0pt,outer sep=0pt] (0,0.15ex) -- (0,1.4ex); }}}}

\newcommand{\outs}[2]{\ensuremath{\ovl #1 (#2)}}

\newcommand{\inps}[2]{\ensuremath{#1 (#2)}}

\newcommand{\outb}[3]{\ensuremath{\ovl #1 (#2, #3)}}
\newcommand{\inpb}[3]{\ensuremath{#1 (#2, #3)}}

\newcommand{\wnouts}[2]{\ensuremath{{{\wn}} \hspace*{1pt} \outs{#1}{#2}}}

\newcommand{\lns}{\ensuremath{\\[0pt]}}

\newcommand{\sN}{\textbf{sN}}

\newcommand{\textpar}[1]{\textit{\textbf{#1}}}

\newcommand{\subt}{\ensuremath{\preccurlyeq}}

\newcommand{\trule}[3]{\ensuremath{ \inferrule[{\kw{(#1)}}]{#2}{#3} }}

\newcommand{\vect}[1]{\widetilde{#1}}

\newcommand{\blk}[1]{\ensuremath{(\, #1 \,)}}

\newcommand{\tvar}[1]{\ensuremath{\mathsf{#1}}}

\newcommand{\tyal}[2]{\ensuremath{\tvar{#1} \mapsto {#2}}}
\newcommand{\tyali}[3]{\ensuremath{\tvar{#1}_{#3} \mapsto {#2}_{#3}}}

\newcommand{\subTP}[2]{\ensuremath{ [ \raisebox{1.4pt}{{\small $ #1 $}}  \! /  \raisebox{-1.0pt}{{\small $#2$}} ] }}

\newcounter{NamesCounter}
\newif\ifshownodes
\pgfkeys{/show nodes/.is if=shownodes}
\pgfkeys{/show nodes=true}
\pgfkeys{/show nodes=false}

\tikzfading[name=fade right,
left color=transparent!0,
right color=transparent!100]

\tikzfading[name=fade left,
right color=transparent!0,
left color=transparent!100]

\tikzfading[name=fade out,
inner color=transparent!50,
outer color=blue]

\tikzset{% 
link/.style={line width=1.2pt, color=blue, rounded corners=0.2cm, line cap=round} , 
crossing axiom/.style={link, double distance=1.2pt, draw=white, double=blue}, %line width=1.2pt, draw=white, double=blue},
cut/.style={ link, color=olive },
crossing cut/.style={crossing axiom, double=olive} , 
inactive cut/.style={dashed}, % no scope... cannot reduce, premises are inactive too 
premise/.style={link, dashed},
box/.style={inner sep=2pt, outer sep=0pt, line width=1.2pt, color=purple, rounded corners=0.2cm}, 
crossing boxlink/.style={box, double distance=1.2pt, draw=white, double=purple}, %line width=1.2pt, draw=white, double=blue},
name box/.style={box,dotted},
helpnodes/.style={color=darkgray!60}
}%

\pgfdeclarelayer{background}
\pgfdeclarelayer{foreground}
\pgfsetlayers{background,main,foreground}

\def\blinklen{1.4cm}
\def\blinkangle{55}

\newdimen\xA
  	\newdimen\xB
  	\newdimen\yA
  	\newdimen\yB
  	\newdimen\maxY
  	\newdimen\minY
  	\newdimen\maxX
  	\newdimen\minX
  	\newdimen\midX
  	
\newdimen\userDimA
\newdimen\userDimB 
\newdimen\userDimC
\newcommand{\AxLink}[8]{%
  	\pgfextractx{\xA}{\pgfpointanchor{#5}{center}} 
  	\pgfextractx{\xB}{\pgfpointanchor{#6}{center}}
  	\pgfextracty{\yA}{\pgfpointanchor{#5}{center}} 
  	\pgfextracty{\yB}{\pgfpointanchor{#6}{center}}
  	\pgfmathsetlength{\maxY}{max(\yA,\yB)}
  	\pgfmathsetlength{\maxX}{max(\xA,\xB)}
  	\pgfmathsetlength{\minX}{min(\xA,\xB)}
  	\pgfmathsetlength\midX{\minX+(\maxX -\minX)/2}  

    \draw[anchor=mid] 
       (#5) +(0,0pt) node[color=black] (Ax-#5-#6-L) {{{\small\ensuremath{#1 \tp \phantom{(}\!\! #2 \phantom{#1 \tp (}\!\!}}}} %}
        (#6) +(0,0pt)  node[color=black] (Ax-#5-#6-R) {{{\small\ensuremath{#3 \tp\phantom{(}\!\! #4\phantom{#3 \tp (}\!\!}}}}  ;
    \coordinate (Ax-#5-#6-M) at ($(\midX,\maxY) + (0cm,0.60cm) + (0cm,#7)$);      
      
  	\draw[link,#8,anchor=mid] 
     (Ax-#5-#6-L) %node [color=black] {{{\small\ensuremath{#2}}}}   
         |- 
    	         (Ax-#5-#6-M)  	% ($(\midX,\maxY) + (0cm,0.60cm) + (0cm,#7)$) 
    			 		-|   (Ax-#5-#6-R)  ; %node[color=black] {{{\small\ensuremath{#4}}}};
         \begin{pgfonlayer}{foreground}
         \ifshownodes 
 	     \path (Ax-#5-#6-M.north) + (0.0cm,0.15cm) node[helpnodes] {{\tiny (Ax-[#5]-[#6])}} ;
 	     \path (Ax-#5-#6-L.north) + (0.2cm,0.07cm) node[helpnodes] {{\tiny [L]}} ;
 	     \path (Ax-#5-#6-R.north) + (0.2cm,0.07cm) node[helpnodes] {{\tiny [R]}} ;
         \else {}
         \fi
         \end{pgfonlayer}
}%

\newcommand{\AxLinkB}[4]{%
  	\pgfextractx{\xA}{\pgfpointanchor{#1}{center}} 
  	\pgfextractx{\xB}{\pgfpointanchor{#2}{center}}
  	\pgfextracty{\yA}{\pgfpointanchor{#1}{center}} 
  	\pgfextracty{\yB}{\pgfpointanchor{#2}{center}}
  	\pgfmathsetlength{\maxY}{max(\yA,\yB)}
  	\pgfmathsetlength{\maxX}{max(\xA,\xB)}
  	\pgfmathsetlength{\minX}{min(\xA,\xB)}
  	\pgfmathsetlength\midX{\minX+(\maxX -\minX)/2}  

    \draw[anchor=mid] 
       (#1) +(0,0pt) node[color=black, minimum height=12pt] (Ax-#1-#2-L) {\phantom{()}} %}
        (#2) +(0,0pt)  node[color=black, minimum height=12pt] (Ax-#1-#2-R) {\phantom{()}}  ;
    \coordinate (Ax-#1-#2-M) at ($(\midX,\maxY) + (0cm,0.60cm) + (0cm,#3)$);      
      
  	\draw[link,#4,anchor=mid] 
     (Ax-#1-#2-L) %node [color=black] {{{\small\ensuremath{#2}}}}   
         |- 
    	         (Ax-#1-#2-M)  	% ($(\midX,\maxY) + (0cm,0.60cm) + (0cm,#7)$) 
    			 		-|   (Ax-#1-#2-R)  ; %node[color=black] {{{\small\ensuremath{#4}}}};
         \begin{pgfonlayer}{foreground}
         \ifshownodes 
 	     \path (Ax-#1-#2-M.north) + (0.0cm,0.15cm) node[helpnodes] {{\tiny (Ax-[#1]-[#2])}} ;
 	     \path (Ax-#1-#2-L.north) + (0.2cm,0.07cm) node[helpnodes] {{\tiny [L]}} ;
 	     \path (Ax-#1-#2-R.north) + (0.2cm,0.07cm) node[helpnodes] {{\tiny [R]}} ;
         \else {}
         \fi
         \end{pgfonlayer}
}%

\newcommand{\CutLink}[4]{%
  	\pgfextractx{\xA}{\pgfpointanchor{#1}{center}} 
  	\pgfextractx{\xB}{\pgfpointanchor{#2}{center}}
  	\pgfextracty{\yA}{\pgfpointanchor{#1}{center}} 
  	\pgfextracty{\yB}{\pgfpointanchor{#2}{center}}
  	\pgfmathsetlength{\minY}{min(\yA,\yB)}
  	\pgfmathsetlength{\maxX}{max(\xA,\xB)}
  	\pgfmathsetlength{\minX}{min(\xA,\xB)}
  	\pgfmathsetlength\midX{\minX + (\maxX -\minX)/2}  

  \path[anchor=mid] 
       (#1) +(0cm,0pt) node (#1-#2-CL) [color=black] {}
       (#2) +(0cm,0pt) node (#1-#2-CR) {};
        
  	\path[cut,anchor=mid,#4,draw] 
      (#1-#2-CL)  %node [color=black] {{{\small\ensuremath{#2}}}}   
         |- 
    		 ($(\midX,\minY) + (0cm,-0.50cm) + (0cm,-#3)$) 
    			 		-|   (#1-#2-CR)  ; %node[color=black] {{{\small\ensuremath{#4}}}};
}%

\newcommand{\OkayLink}[6]{
    \path  (#5) +(0,\blinklen) coordinate (#5-Above); 
    \path  (#5) +(\blinkangle:\blinklen) coordinate (#5-RightUp); 
    \path  ($(#5)!(#5-RightUp)!(#5-Above)$) node[minimum height=10pt] (#5-T) {};
     
     \Tnode{#1}{#3}{#5}{#6} ;
     \Tnode{#2}{#4}{#5-T}{#6-T} ;
         
     \path[draw, link, anchor=mid] (#6) -- (#5-T) node[anchor=mid, color=black] {}; 
     
     \pgfextracty{\userDimA}{\pgfpointanchor{#5-T}{center}}     
     \pgfextracty{\userDimB}{\pgfpointanchor{#5}{center}}
     \pgfmathsetlength\userDimC{(\userDimA-\userDimB)/2}  
     
     \path (#5) +(0,\userDimC + 1.5pt) coordinate (#6-at-crd); 
     \begin{pgfonlayer}{foreground}
       \path (#6-at-crd) node[draw, box, rectangle, rounded corners=2pt, minimum width=10pt, minimum height=4pt,fill=white, text=black] 
          (#6-at) {{\small\ensuremath{\sat}}};
     \end{pgfonlayer} 
}

\newcommand{\UnaLink}[6]{
    \path  (#5) +(0,\blinklen) coordinate (#5-Above); 
    \path  (#5) +(\blinkangle:\blinklen) coordinate (#5-RightUp); 
    \path  ($(#5)!(#5-RightUp)!(#5-Above)$) node[minimum height=10pt] (#5-T) {};
     
     \Tnode{#1}{#3}{#5}{#6} ;
     \Tnode{#2}{#4}{#5-T}{#6-T} ;
         
     \path[draw, link, anchor=mid] (#6) -- (#5-T) node[anchor=mid, color=black] {}; 
}

\newcommand{\BinLink}[6]{%
    \path (#5) +(180-\blinkangle:\blinklen)  node[minimum height=10pt] (#6-L) {};	
	\path (#5) +(\blinkangle:\blinklen)  node[minimum height=10pt] (#6-R) {}; % {{\small\ensuremath{#4}}}; 
	 \path[draw,anchor=mid]
	 (#5) +(0cm,0.0pt) node[left] {\small \ensuremath{#1 \tp #2\phantom{(}\!}}   
	 (#5) +(0cm,0.0pt) node[right] {\small\ensuremath{\!\phantom{(}#4}}
      (#5) +(0cm,0.0pt) node(#6) 
        [ %anchor=mid, 
         ] 
		{\small \ensuremath{#3}}  
        [link] 
        (#6) -- (#6-R) %($(\pD) + (0cm,0.1cm)$) 
        node[anchor=mid, color=black] {} % {{\small\ensuremath{#4}}} 
        (#6) -- (#6-L) %($(pC) + (0cm,2.4pt)$) 
         node[anchor=mid, color=black] {}; % {{\small\ensuremath{#2}}};
         \ifshownodes 
 	     \path (#6.north east) + (0.25cm,0.07cm) node[helpnodes] {{\tiny (#6)}} ;
         \else {}
         \fi
}%

\newcommand{\Tnode}[4]{%
	\draw[anchor=mid,color=black] %, inner sep=2.4pt] 
      (#3) +(0cm,0.0pt) node[anchor=mid,color=black] (#4) {{{\small\ensuremath{#1 \tp \phantom{(}\!\! #2 \phantom{#1 \tp (}\!\!}}}} ;
    \ifshownodes 
 	   \path (#4.north) + (0.55cm,0cm) node[helpnodes] {{\tiny (#4)}} ;
     \else {}
    \fi
}%

\newcommand{\Snode}[3]{%
	\draw[anchor=mid,color=black] %, inner sep=2.4pt] 
      (#2) +(0cm,0.0pt) node[anchor=mid,color=black] (#3) {\ensuremath{#1}} ;
    \ifshownodes 
 	   \path (#3.north) + (0.55cm,0cm) node[helpnodes] {{\tiny (#3)}} ;
     \else {}
    \fi
}%

\newcommand{\aFittedBox}[3]{%
\begin{pgfonlayer}{background}
    	\node[draw,box, fit=#2,inner sep=6pt,#3] (#1) {}; 
\end{pgfonlayer}
\ifshownodes 
 	\path (#1.north east) + (0.1cm,0.15cm) node[helpnodes] {{\tiny (#1)}} ;
\else {}
\fi
}%

\newdimen\tSize
\newdimen\tPos

\newcommand{\RecVarLink}[6][1.2cm]{ 

  \pgfmathsetlength{\tSize}{0pt}
  \pgfmathsetlength{\tPos}{0pt}
    \foreach \name / \type / \len [count=\j] in {#6}
           { 
	              \pgfmathaddtolength\tSize{\len}
	              \global\tSize=\tSize 
           };
           
    \path 
    (#2) node[draw, box, rectangle, rounded corners=2pt, minimum width=14pt, minimum height=12pt,fill=white,text=black] 
       (#3) {{\small\ensuremath{#4}}};
    
    \ifshownodes 
 	   \path (#3.east) + (0.35cm,0.0cm) node[helpnodes] {{\tiny (#3)}} ;
     \else {}
    \fi    

   \path 
         let \p1=(#2) in   let \n1={\tSize/2}       in 
                  (#2) +(-\n1,-#1) coordinate (#3-init)  ;
   
   \path (#3-init) coordinate (#3-a-1);  
           
  	 \foreach \name / \type / \dist [count=\i] in {#6}  
  	  {   
  	 	  \ifnum \i>1 
      	 		\path ($ (#3-init) + (\tPos,0) $) coordinate (#3-a-\i) ;
           \fi
           
          \pgfmathaddtolength{\tPos}{\dist}
           \global\tPos=\tPos 
 	  }    
    
    \foreach \name / \type / \dist [count=\i] in {#6}  
        { 
           \Tnode{\name}{\type}{#3-a-\i}{#3-c-\i} ;  % add Tnode
           \begin{pgfonlayer}{background}
    		   \draw[link] (#3.south) -| (#3-c-\i) {};   % connect origin node to this one 
            \end{pgfonlayer}  
    	    }
}%
\newcommand{\RecLink}[6][1.2cm]{%

  \pgfmathsetlength{\tSize}{0pt}
  \pgfmathsetlength{\tPos}{0pt}
 
    \foreach \var / \name / \type / \len [count=\j] in {#6}
           { 
	              \pgfmathaddtolength\tSize{\len}
	              \global\tSize=\tSize 
           }; 
 
    \path (#2) node[draw, box, rectangle, rounded corners=2pt, minimum width=14pt, minimum height=12pt,fill=white, text=black] 
       (#3) {{\small\ensuremath{#4}}}; % {}; 
    \ifshownodes 
 	   \path (#3.east) + (0.35cm,0.0cm) node[helpnodes] {{\tiny (#3)}} ;
     \else {}
    \fi    
    
    \path 
         let \p1=(#2) in   let \n1={\tSize/2}       in 
                  (#2) +(-\n1,-#1) coordinate (#3-c-init)  
                  (#2) +(-\n1,#1) coordinate (#3-x-init) ;
   
   \path (#3-c-init) coordinate (#3-a-1);  
   \path (#3-x-init) coordinate (#3-b-1);
           
  	 \foreach \var / \name / \type / \dist [count=\i] in {#6}  
  	  {   
  	 	  \ifnum \i>1 
      	 		\path ($ (#3-c-init) + (\tPos,0) $) coordinate (#3-a-\i) ;
      	 		\path ($ (#3-x-init) + (\tPos,0) $) coordinate (#3-b-\i) ;
           \fi
           
          \pgfmathaddtolength{\tPos}{\dist}
           \global\tPos=\tPos 
 	  }    
    
    \foreach \var / \name / \type / \dist [count=\i] in {#6}  
        { 
           \Tnode{\name}{\type}{#3-a-\i}{#3-c-\i} ;  % add Tnode
           \Tnode{\var}{\type}{#3-b-\i}{#3-x-\i} ;           
           
           \begin{pgfonlayer}{background}
    		   \draw[link] (#3.south) -| (#3-c-\i) {};   % connect origin node to this one 
    		   \draw[link] (#3.north) -| (#3-x-\i) {}; 
            \end{pgfonlayer}  
    	    }
}%
\newcommand{\RecLinkB}[6][1.2cm]{%

    \path  (#2) +(0,\blinklen) coordinate (#2-Above); 
    \path  (#2) +(\blinkangle:\blinklen) coordinate (#2-RightUp); 
    \path  ($(#2)!(#2-RightUp)!(#2-Above)$) node[minimum height=10pt] (#2-T) {};

    \pgfextracty{\userDimA}{\pgfpointanchor{#2-T}{center}}     
    \pgfextracty{\userDimB}{\pgfpointanchor{#2}{center}}
    \pgfmathsetlength\userDimC{(\userDimA-\userDimB)/2}  %% mu blob here - new origin 
 
    \path (#2) +(0,\userDimC) coordinate (#2-c-mu);

    \pgfmathsetlength{\tSize}{0pt}
    \pgfmathsetlength{\tPos}{0pt}
 
    \foreach \var / \name / \type / \len [count=\j] in {#6}
           { 
	              \pgfmathaddtolength\tSize{\len}
	              \global\tSize=\tSize 
           }; 
 
    \path (#2-c-mu) 
    node[draw, box, rectangle, rounded corners=2pt, minimum width=14pt, minimum height=12pt,fill=white, text=black] 
       (#3) {{\small\ensuremath{#4}}}; % {}; 
    \ifshownodes 
 	   \path (#3.east) + (0.35cm,0.0cm) node[helpnodes] {{\tiny (#3)}} ;
     \else {}
    \fi    
    
    \path 
         let \p1=(#2-c-mu) in   let \n1={\tSize/2}       in 
                  (#2-c-mu) +(-\n1,-\userDimC) coordinate (#3-c-init)  
                  (#2-c-mu) +(-\n1,\userDimC) coordinate (#3-x-init) ;
   
   \path (#3-c-init) coordinate (#3-a-1);  
   \path (#3-x-init) coordinate (#3-b-1);
           
  	 \foreach \var / \name / \type / \dist [count=\i] in {#6}  
  	  {   
  	 	  \ifnum \i>1 
      	 		\path ($ (#3-c-init) + (\tPos,0) $) coordinate (#3-a-\i) ;
      	 		\path ($ (#3-x-init) + (\tPos,0) $) coordinate (#3-b-\i) ;
           \fi
           
          \pgfmathaddtolength{\tPos}{\dist}
           \global\tPos=\tPos 
 	  }    
    
    \foreach \var / \name / \type / \dist [count=\i] in {#6}  
        { 
           \Tnode{\name}{\type}{#3-a-\i}{#3-c-\i} ;  % add Tnode
           \Tnode{\var}{\type}{#3-b-\i}{#3-x-\i} ;           
           
           \begin{pgfonlayer}{background}
    		   \draw[link] (#3.south) -| (#3-c-\i) {};   % connect origin node to this one 
    		   \draw[link] (#3.north) -| (#3-x-\i) {}; 
            \end{pgfonlayer}  
    	    } ;
}%

\newcommand{\mypar}[1]{\vspace{3.4pt}\noindent {\textpar{#1}}}
\title{Multiparty Sessions based on Proof Nets}
\author{Dimitris Mostrous
\institute{LaSIGE, Department of Informatics,  Faculty of Engineering\\
University of Lisbon, Portugal.} 
\email{dimitris@di.fc.ul.pt} 
}
\def\titlerunning{Multiparty Sessions based on Proof Nets}
\def\authorrunning{D. Mostrous}

\begin{document} %\selectlanguage{english}
\maketitle

\begin{abstract}
We interpret Linear Logic Proof Nets in a term language based on Solos calculus. 
The system includes a synchronisation mechanism, obtained by a conservative extension of the logic, 
that enables to define non-deterministic behaviours and multiparty sessions.  
\end{abstract}

\section{Introduction}\label{sec:intro} 

%
%With the emergence of manycore computing our notion of evaluation strategy 
%needs to be as flexible as possible. 
%
%The programs-as-proofs approach to structured interactions, 
%initiated by the recent work~\cite{Caires:2010:SILL} 
%on Intuitionistic \LL and session types, has renewed interest  
%
%starting from has renewed interest in 
%logical systems for structured interactions. 
%
%Parallelism is not incompatible with structured communications. 
%
%What is the purpose of imposing an ordering of reductions in a typed process algebra that is confluent and strongly normalising? 
%From the rather obvious answer to this question, one discovers that 
%an accurate interpretation of 
%Linear Logic\footnote{Let us say, without commutative reductions which are known to cause non-confluence, although semantically the argument still holds~\cite{Girard87}.} 
%cannot benefit from using prefixed communications unless absolutely necessary, 
%which may seem to contradict what we are accustomed to 
%use in the (highly non-deterministic) \pic !

Since their inception, sessions~\cite{honda.vasconcelos.kubo:language-primitives,THK} 
and multiparty sessions~\cite{HondaK:mulast} have been gaining 
momentum as a very useful foundation for the description and verification 
of {\em structured interactions}. 
Interestingly, recent works have established a close correspondence between 
typed, synchronous pi-calculus processes and sequent proofs of 
a variation of Intuitionistic Linear Logic~\cite{Caires:2010:SILL}. 
This particular interpretation of Linear proofs is 
considered a sessions system because it has (for practical purposes) the same 
type constructors but with a clear logical motivation. 
%because it facilitates the description 
%and verification of structured interactions based on message 
%passing. %, using familiar sessions primitives. 
%Indeed, session types are closely related to formulae of Linear Logic. 
  In this paper we outline a system based on an interpretation of the proof objects of 
Classical Linear Logic, namely Proof Nets~\cite{Girard87}, improving our previous work~\cite{mostrous12}. 
The process language 
resembles Solos~\cite{laneve.victor:solos-concert} and exhibits asynchrony 
in both input and output. Proof Nets have a number of advantages 
over sequent proofs, such as increased potential for parallelism 
and a very appealing graphical notation that could be seen as a new kind of 
{\em global type}~\cite{HondaK:mulast}.  
%We claim that this interpretation is more canonical and simple, since 
%it is founded on the canonical proof objects of the logic. 
  Nevertheless, accurate logical interpretations are typically 
deterministic, which limits their applicability to concurrent 
programming. However, with a very modest adaptation that enables 
{\em synchronisation}, non-deterministic behaviours can be 
allowed without compromising the basic properties of interest, %linear proofs, 
namely strong normalisation and deadlock-freedom. 

Let us distinguish \textit{multiparty behaviours} and the 
\textit{multiparty session types} (global types) of~\cite{HondaK:mulast}.
A multiparty behaviour emerges when more than two processes 
can be part of the same session, and this is achieved at the operational 
level by a synchronisation mechanism such as the multicast  
request $\ovl a[\tsf{2..n}](\vec{s}).P$~\cite{HondaK:mulast}. 
We propose a similar mechanism in the form of replications with synchronisation, 
$\oc a_1(x_1) \cdots a_n(x_n).P$, which allow a service to be activated 
with multiple parties. 
A global type captures the interactions and sequencing constraints of 
the complete protocol of a program. 
In our proposal, the equivalent to a global type is the {\em proof net} 
of the program. 
Although our approach is technically very different, we believe 
that the logical foundations and simpler meta-theory 
are appealing. 
We show how pi-calculus channels with \tsf{i/o} type and 
a multi-party 
interaction from~\cite{HondaK:mulast} can be encoded. 

%%
%% sN is very weak in the lazy form: it allows !a?b | !b?a
%%
%
%\begin{enumerate}
%\item Put at least one \PN\ reduction diagrammatically.
%\item Put an actual multi-party example (from POPL). 
%\item Explain that in a many-core setting, we want as much parallelism as possible. 
%  This is combined with the fact that linear and exponential reductions commute, and 
%  therefore there is really \textbf{no point} into having a linear prefix. 
%\item Explain tha m-party sessions have complex semantics.
%\item explain that m-party types have weaker properties and a more complex meta-theory. 
%  POPL has 14 reduction rules. 
%\item Projection is a tool, not the aim. 
%\item Explain that due to confluence, linear prefixes cannot alter the result 
%  or have any effect. 
%\end{enumerate}
%

\section{The Process Interpretation}\label{sec:system}

\mypar{Syntax} 
%Our language is similar to the Solos calculus~\cite{laneve.victor:solos-concert}.  
%We also include the usual form of branching and selection from sessions~\cite{honda.vasconcelos.kubo:language-primitives} 
%and a form of explicit substitutions,   
%as well as replication, which is typically encoded in untyped solos.
%Of course, all these constructs 
The language is inspired by proof nets except that connectives have explicit 
locations (names). %, 
%from which we recovered this polymorphic solos-like language. 
Types %, defined later, 
are ranged over by $A, B, C$, with type variables ranging over \tvar X, \tvar Y. 
We assume a countable set of {\em names}, ranged over by $a, b, c, x, y, z ,r, k$.    
Then, $\vect b$ stands for a sequence $b_1,\ldots,b_n$  
of length $|\vect b| = n$, and similarly for types. 
Processes, %ranged over by 
$P, Q, R$, are defined as follows:% 
\noindent {\small \begin{gather*}
\begin{array}{rcllllll}
P & ::=  & \outs a {\vect A, \vect x}  
			&		\bprl \:\: \inps a {\vect{\tvar X}, \vect y} 
			&		\bprl \:\: \choiceI a i b 
			&		\bprl \:\: \branchIT a i x A P I 
			&		\bprl \:\: \wnouts {a} {b}
   			&	    \bprl \:\: \oc a_i(x_i \tps A_i)_{i \in I}.P
\\[4pt]
   &  \bprl &  ab 
   	       & \bprl \:\: \tyal X A 
   	       & \bprl \:\: \newc{a \tps A} P 
   	       & \bprl \:\: \newc{\tvar X} P 
   	       & \bprl \:\: (P \prls Q)  
   	       & \bprl \:\: \nil   \\
\end{array}
\end{gather*}}%
%
%
%\noindent {\small \begin{gather*}%    %% \outs a {\vect A, \vect x} \prl \inps a {\vect{\tvar X}, \vect y}  
%\begin{array}{ccccccl}
%P  &   ::=  &  \outs a {\vect A, \vect x}  & \bprl & \inps a {\vect{\tvar X}, \vect y}            && \textit{(output, input)}
%%  &  \outs a {\vect v}  & \bprl & \inpv a v           && \textit{(output, input)} 
%\\[4pt]
%   &  \bprl  & \choiceI a i b            & \bprl & \branchIT a i x A P I && \textit{(select, branch)}
%\\[4pt]
%    & \bprl & ab             & \bprl & \tyal X A         && \textit{(substitution, type alias)} 
%\\[4pt]
%    & \bprl & 
%           \wnouts {a} {b} 
%                                    & \bprl & \oc a_i(x_i \tps A_i)_{i \in I}.P %\oc a_1(x_1 \tps A_1) \cdots a_n(x_n \tps A_n).P       
%                                        \hspace*{-10pt} & & \textit{(request, accept)} 
%\\[4pt]
%    & \bprl & \newc{a \tps A} P            
%                                    & \bprl & \newc{\tvar X} P && \textit{(new name, type variable)} 
%\\[4pt]
%    & \bprl & (P \prls Q)  & \bprl & \nil                   && \textit{(parallel, inaction)} 
%\end{array}
%%\\[8pt]
%%\begin{array}{ccccccc}   %% v \, ::= \, a \prl \kil \prl \btvar X \prl A
%%v  & ::=  &   a                    & \bprl & A 
%%\\
%%    &       & \textit{(name)} &            &  \textit{(type)}
%%\end{array}
%\end{gather*}}%
%
%  ~\cite{Pierce:2000:BEP:long} 
%
There are two kinds of {\em solos}-like~\cite{laneve.victor:solos-concert} communication devices: 
$\inps a {\vect{\tvar X}, \vect y}$ %which can be understood as an input, 
and  $\outs a {\vect A, \vect x}$. %which can be seen as an output, although the distinction is  %%$\outv a v$ 
%not substantial in the absence of prefix.  
In typed processes, we will be using 
the nullary signals  
$a$ for $a()$ and $\ovl a$ for 
$\outs a {}$,\footnote{They have no computational content, but without them reduction leaves garbage axioms of unit type.} 
the binary input $\inpb a b c$ (resp. output $\outb a b c$) 
and the {\em asynchronous} polymorphic input $\inpb a {\tvar X} b$ (resp. output $\outb a B b$). 
The {explicit substitution}, $ab$, interprets \LL\ axioms; this is standard in related works~\cite{mostrous12,Caires:esop13}. %, 
%with some corrections in the dynamics to avoid a soundness problem in the presence of asynchrony.  
The type alias $\tyal X A$ is simply a typing device,  %, it never results to an actual substitution $\subTP{A\,}{\tvar X}$ during reduction.  
%The type variable $\tvar X$ is free, except under $\newc{\tvar X} P$.   
and the scope of $\tvar X$ is restricted with $\newc{\tvar X} P$. 
The branching connective $\branchIT a i x A P I$, with $I = \set{1,\dotsc, n}$, written also in the form    
$\branch a { \methI 1 {x_1 \tps A_1} {P_1} \orBra \cdots \orBra \methI n {x_n \tps A_n} {P_n} }$, 
offers an %ordered 
indexed sequence of alternative behaviours.  
One of these can be selected using $\choiceI a k b$ with $k \in I$. 
%To ease the presentation of examples, we will sometimes 
%use a more readable notation, for example $\branch a { \methK {read} x P \orBra \methK {write} y Q }$, 
%and similarly $\choice a {read} b$.
%%\footnote{% 
%%For an unambiguous correspondence with indices, we assume 
%%that labels in such examples will be strictly ordered using the lexicographic ordering. 
%%The reason we do not use {\em labeled branching} is that 
%%we want to avoid the introduction of labels in the logical type structure, 
%%and the solution is equivalent in every respect.}% 
%
Our %novel 
notion of replication enables synchronisation, similarly to a multiparty \qt{accept} (cf.~\cite{HondaK:mulast}). 
The notation is $\oc a_i(x_i \tps A_i)_{i \in I}.P$, written also as $\oc a_1(x_1 \tps A_1) \cdots a_n(x_n \tps A_n).P$    
with $n \geq 1$. 
%To invoke a replication, we use
Dually, $\wnouts a {b}$ %, which 
can be thought as a \qt{request.}

%In the following we sometimes omit the types to simplify the presentation. 

%\begin{rema}[On the overlap of features]\label{rema:design}
%As in~\cite{Pierce:2000:BEP:long}, where communications are combined with 
%second-order elements, our purpose is to obtain small, concise calculus. 
%A programming language would probably define  
%more constructors such that these features %(second-order, name fusions, \etc)  
%are orthogonal. We believe that, nevertheless, no confusion arises from  
%the current presentation, and  we find it convenient to work in this setting. 
%\end{rema}

%

\mypar{Free, passive, and active names}  
The {\em free names} ($\fn P$) are defined in the standard way. We just note 
that the only bound names are $a$ in $\newc{a\tps A}\,P$ and  
the $x_i$ in $\branchIT a i x A P I$ and $\oc a_i(x_i \tps A_i)_{i \in I}.P$. 
The {\em passive names} ($\pn P$) are defined similarly to \fn P except for: 
{\small \begin{gather*}
\pn{\inps a {\vect{\tvar X}, \vect x}} = \pn{\outs a {\vect A, \vect x}} = \set{\vect x} 
\qquad 
\pn{ \choiceI a i b } = \pn{\wnouts a b} = \set{b}   
\qquad
\pn{ab} = \emptyset  
%\qquad
%\pn{\newc{a\tps A} P} = \pn{P} \setminus a
\\
\pn{ \branchIT a i x A P I } =  \cup_{i\in I} ( \pn{ P_i } \setminus \set{x_i} ) 
\qquad
\pn{ \oc a_i(x_i \tps A_i)_{i \in I}.P } = \pn{P} \setminus \cup_{i\in I} \set{x_i} 
\end{gather*}}%  
The {\em active names} ($\an P$) are defined by $\an P = \fn P \setminus \pn P$. 
% 
%If $a \in \an P$ then $a$ appears only as conclusion in $P$;  
%and this is how we avoid \qt{touching} premises in reduction. 
For example, $y$ in $\newc{x} \blk{\inps a {x,y} \prls xy}$ is not active.
(As usual, we assume the name convention.) %~\cite{Bar84}.)
%
% keep, it's concise..
%
%For example, in $\newc{x} \blk{ bx \prls \inps{a}{x,y} \prls yc }$ we cannot 
%apply the substitution $bx$, because the conclusion $b$ would become a premise: 
%$\inps{a}{b,y} \prls yc$. 

%This easily leads to loss of type soundness as we explain later.    

%
%\paragraph{\textpar{Free type variables}} 
%The free type variables (\ftv P) are defined in a standard way, noting that % $\ftv{\btvar X} \defeq \tvar X$ 
%$\ftv{\newc{\tvar X} P} = \ftv{P}\setminus\tvar X$. 
%The free type variables of a type $A$ (\ftv A) are defined later. 

\mypar{Structure Equivalence}
With $\equiv$ we denote the least congruence on processes that is an equivalence relation, 
equates processes up to \aconv, 
satisfies the abelian monoid laws for parallel composition, the usual laws for scope 
extrusion, %(including for new type variables), 
and satisfies the following 
axioms:\footnote{The free type variables (\ftv P) are defined in a standard way, noting that 
$\ftv{\newc{\tvar X} P} = \ftv{P}\setminus\tvar X$. 
The free type variables of a type $A$ (\ftv A) are also standard and arise from $\forall$/$\exists$.}
{\small \begin{gather*}
%(P \prl Q) \prl R \equiv P \prl (Q \prl R) 
%\qquad
%P \prl Q \equiv Q \prl P 
%\\
%P \prl \nil \equiv P 
%\qquad
%\newc{a\tps A} \nil \equiv \nil 
%\qquad
\newc{\tvar X} \nil \equiv \nil 
%\\
%\newc{a\tps A} P \prl Q \equiv \newc{a\tps A} \blk{P \prl Q} \quad (a\not\in\fn Q) 
%\qquad
%\newc{a\tps A}\newc{b\tps B} P \equiv \newc{b\tps B}\newc{a\tps A} P
%%\quad 
%%%\newc{a\tp A}\newc{b\tp B} ab.P \equiv \nil            % old
%%\newc{a\tp A} ab.P \equiv \nil \quad (a\not\in\fn P) 
%\\
\qquad 
\newc{\tvar X} P \prl Q \equiv \newc{\tvar X} \blk{P \prl Q} \quad (\tvar X\not\in\ftv Q) 
\\
\newc{\tvar X}\newc{\tvar Y} P \equiv \newc{\tvar Y}\newc{\tvar X} P
\qquad 
\newc{\tvar X}\newc{a \tps A} P \equiv \newc{a \tps A}\newc{\tvar X} P \quad (\tvar X\not\in\ftv A)
\\ 
ab\equiv ba  
\qquad
ab \prl \oc a_1(x_1 \tps A_1) \cdots a(x \tps A) \cdots a_n(x_n \tps A_n).P
\equiv 
ab \prl \oc a_1(x_1 \tps A_1) \cdots b(x \tps A) \cdots a_n(x_n \tps A_n).P
\end{gather*}}%
The most notable axiom is the last one, which effects a forwarding, \eg, 
$\wnouts a {x} \prl ab \prl \oc b(y) . P \equiv^2 ab \prl \wnouts a {x} \prl \oc a(y) . P$. 
%$ \R ab \prl P\subs x y \prl \oc a(y) . P$. 
As can be seen next, the term on the right can now reduce. %, 
%Simply put, axioms with exponential type allow forwarding without reduction, 
%while in related works a $\newc{b}$ would be needed, and this would 
%effectively fix the number of clients to $b$. 
%
%Notice how %: %in the penultimate line;  
%$\oc$-names can change following a path formed by axioms,  
%%, 
%%effecting a delayed substitution (locally) without actually applying it.  
%%%The result is a kind of open reduction for exponentials, which means that  
%%%not all cuts need to be under a name restriction. 
%overcoming a limitation of related works, where 
%communications must be under restriction (in this case \newc{a} or \newc{b}). 
%%Concretely, this means that terms that invoke a server can be added 
%%dynamically, which is impossible with existing interpretations.   
%%%This is without doubt an improvement 
%%%compared to other interpretations.  

%
\mypar{Reduction} \qt{$\R$} is the smallest binary relation 
on terms such that: 
%
%      &  \bprl  & \choiceI a i b            & \bprl & \branchIT a i x A P I && \textit{(select, branch)}
%
{\small \begin{gather*}
\begin{array}{rclrr}
\outs a {\vect A, \vect y} \prl \inps a {\vect{\tvar X}, \vect x}  
   & \R & \vect{\tyal X A} \prl \vect{xy} & \quad |\vect A| = |\vect{\tvar X}|,\: |\vect x| = |\vect y|  & \tsf{(R-Com)}
 \\[4pt]
 \choiceI a k b \prl \branchIT a i x A P I & \R & P_k\subs b {x_k} & k \in I & \tsf{(R-Sel)}
 \\[4pt]
%\wnouts{a_1}{b_1} \prls \cdots \prls \wnouts{a_n}{b_n} 
%\prl  \oc a_1(x_1 \tps A_1) \cdots a_n(x_n \tps A_n).P
%        & \R & 
%         P \subs{b_1,\ldots,b_n}{x_1,\ldots,x_n} \prl \oc a_1(x_1 \tps A_1) \cdots a_n(x_n \tps A_n).P
\prod_{i \in I} \wnouts{a_i}{b_i} 
\prl  \oc a_i(x_i \tps A_i)_{i \in I}.P
        & \R & 
        \multicolumn{2}{l}{ 
         P \subs{b_i}{x_i}_{i\in I} \prl \oc a_i(x_i \tps A_i)_{i \in I}.P
         } \hspace*{30pt} 
           & \tsf{(R-Sync)}
\\[4pt]
\newc{a\tps A} \blk{ ab \prl P } & \R & 
   P \subs{b}{a}  & \hspace*{-15pt}  a \not = b,\:  a \in \an P   & \tsf{(R-Ax)} 
\\[4pt]
\end{array}
\\
P \equiv P' \Rq Q' \equiv Q \:\:\Rightarrow\:\: P \Rq Q \quad \tsf{(R-Str)}
\qquad
P \Rq Q \:\:\Rightarrow\:\: \ctx P \Rq \ctx Q \quad \tsf{(R-Ctx)} 
\\[4pt]
\begin{array}{rcc}
\textit{Contexts in } \tsf{(R-Ctx)}: & \multicolumn{2}{l}{ \quad \ctx{ \cdot } 
\: ::= \:  \cdot \:\prl\: \blk{ \ctx{\cdot} \prls P} \:\prl\: \newc{a \tp A}\, \ctx\cdot  
 \:\prl\: \newc{\tvar X}\, \ctx\cdot }
\end{array}
\end{gather*}}%
%

%\paragraph{\textpar{Reduction rules}}
\tsf{(R-Com)} 
%implements a fusion of $\vect{v}$ and $\vect{u}$ as described above.  
%For names, it 
resembles solos reduction~\cite{laneve.victor:solos-concert} but with explicit fusions~\cite{GardnerLW07,mostrous12}. 
Specifically, given two vectors $\vect x$ and $\vect y$ of length $n$, the notation $\vect{xy}$ stands for 
$x_1y_1 \prls \cdots \prls x_ny_n$ or \nil\ if the vectors 
are empty. 
For polymorphism we create type aliases: 
$\vect{\tyal X A}$ stands for $\tyali X A 1 \prls \cdots \prls \tyali X A n$.  
Combined type and name communication appears also in a synchronous setting~\cite{Pierce:2000:BEP:long}. 
%, although here 
%it is implemented completely asynchronously. 
%Typically, there will be zero or one type elements.
%The second-order (polymorphic) fragment is also based on a new kind of fusion that creates the type aliases $\tyali X A i$. 
%, 
%avoiding the need for prefixes and for type substitution.  
%The reason for which $\vect e$ is ignored will be evident once 
%the typing system is introduced, but we can say that in typed processes $\vect e$ will always be empty 
%when this specific reduction is applied.  
\tsf{(R-Sel)} is standard. 
%
%
%The presence of the scope $\newc{a}(\cdots)$ in \kw{(R2)} implies that $a$ is {\em linear}.  
%%reduces non-determinism, allowing a limited identification of 
%%input and output when they occur linearly. 
%Without this requirement,   
%$a(z) \prls a(k) \prls \rdeft t x B c a y A P$ would have a redex  
%$a(z) \prls a(k)$, which is too unconstrained for disciplined concurrency.  
%In $A\pi$ such a situation is avoided by the distinction of input-output polarities $\ovl a, a$;  
%in our case \kw{(R1)} and \kw{(R2)} are mutually exclusive by {\em linearity}. 
%
% 
In \tsf{(R-Sync)} we synchronise on 
all $a_i$, obtaining a form of multi-party session against $\wnouts {a_1} {b_1} \prls \cdots \prls \wnouts {a_n} {b_n}$. 
\tsf{(R-Ax)} effects a capture-avoiding name substitution,  % and releases the suspended term $P$. 
defined in the standard way.
%one of $a, b$ %does not appear in passive position. 
%appears 
%%at least once 
%in active position. 
The side-condition $a \not = b$ guarantees that no bound name becomes free;  % premises are never identified:  
$a \in \an P$ ensures that the cut is applied correctly, that is, against two (or more) conclusions. 
%
%The other rules are standard. 
%We do not allow reduction under prefix, but it would be safe. %, such as $ab.P$, 

%\begin{rema}[
\mypar{The Caires-Pfenning axiom reduction} %]\label{rema:cfax}
%
%
%We compare briefly our notion of \qt{axiom cut} and the one 
%in~\cite{deyoung_et_al:LIPIcs} 
%which 
%follows~\cite{PerezCPT12}.%,Caires:esop13,Caires2013}.
%\footnote{The reduction rule is not 
%mentioned in~\cite{deyoung_et_al:LIPIcs}, but the type rule is given and 
%one of the authors relayed to me that reduction is assumed to be the same as in~\cite{PerezCPT12}.}
%in~\cite{Abramsky93CILL,PerezCPT12,Caires:esop13,Caires2013}. 
%
%First of all, we implement the axiom cut in a very particular way: 
%it has a continuation, which is original, and it has t
%In particular we focus on our side-condition $a \in \san P$  
%which does not appear in the aforementioned works. 
%(Abramsky uses an {\em active name} condition, but it serves a different purpose.)
%
\tsf{(R-Ax)} is based on 
$\newc{a}\blk{ [a \leftrightarrow b] \prl P } \R P\subs b a \: (a \not= b)$ from~\cite{PerezCPT12},
% works we find the rule: 
%, % 
%
which is similar to the \qt{Cleanup} rule of~\cite{Abramsky93CILL}.
However, in an asynchronous language, this rule breaks subject reduction, 
which motivates our side-condition $a \in \an P$. % is necessary (in one form or another) 
%for type soundness. 
For example,  
$Q \defeq \newc{x,y} \blk{ \ovl a \out{x,y} \prls [x \leftrightarrow b] \prls [y \leftrightarrow c] }$ 
is typable in the system of~\cite{deyoung_et_al:LIPIcs}, 
with conclusion $b\tp A, c \tp B \vdash Q \, :: \, a\tp A \otimes B$, 
but it reduces to   
$\newc{y} \blk{ \ovl a \out{b,y} \prls [y \leftrightarrow c] }$ which is not typable.% any more since  
%a premise is missing.%After one more step we obtain $\ovl a \out{b,c}$ which has no premises at all. 
%In both cases, subject reduction fails. 
%
%The reduction rule is not mentioned in~\cite{deyoung_et_al:LIPIcs}, but the type rule is given and 
%the authors relayed to me that it is assumed to be the same as in~\cite{PerezCPT12}.
%
\footnote{The reduction rule is not 
mentioned in~\cite{deyoung_et_al:LIPIcs}, but the type rule is given and 
one of the authors relayed to me that reduction is assumed to be the same as in~\cite{PerezCPT12}.} 

\mypar{Types and duality} % 
The types, ranged over by $A, B, C, D$\ldots,  %, which 
are linear logic formulae~\cite{Girard87}: %, but without atoms 
%and additive units:% 
{\small \begin{gather*}
A \: ::= \: \munit \bprl \bot \bprl A \otimes B \bprl A \parr B \bprl A \with B \bprl A \oplus B  \bprl \ocm A \bprl \wnm A 
  \bprl \forall\tvar{X}.A \bprl \exists\tvar{X}.A 
 \bprl \tvar X \bprl \dual{\tvar X}
\end{gather*}}%
The {\em mode} \md\ can be \mde\ (empty) or \mdi\ (synchronising):  
\mde\ is a formality and is never shown; \mdi\ is used to enforce some restrictions, 
but does not generally alter the meaning of types.
%\paragraph{\textpar{Negation (Duality)}} 
Negation $\dual{(\,\cdot\,)}$, which corresponds to duality, 
is an involution on types ($\dual{(\dual A)} = A$) defined in the usual way 
(we use the notation from~\cite{TBS}):% 
{\small \begin{gather*}
\dual \munit  \defeq  \bot  \quad\quad 
\dual \bot \defeq \munit  \quad\quad 
\dual{(A \otimes B)} \defeq  \dual A \parr \dual B \quad\quad
\dual{(A \parr B)} \defeq \dual A \otimes \dual B 
\\
\dual{(A \with B)} \defeq \dual A \oplus \dual B \quad\quad 
\dual{(A \oplus B)} \defeq \dual A \with \dual B \quad\quad 
\dual{(\ocm {A})} \defeq \wnm {\dual A} \quad\quad 
\dual{(\wnm {A})}  \defeq \ocm {\dual A} \
\\ 
\dual{(\forall\tvar{X}.A)} \defeq \exists\tvar{X}.\dual A  \quad\quad 
\dual{(\exists\tvar{X}.A)} \defeq \forall\tvar{X}.\dual A \quad\quad 
\dual{(\dual{\tvar{X}})} \defeq \tvar{X}
\end{gather*}}%
%
%
%
%{\small \begin{gather*} %\small
%%
%\begin{array}{rclcrcl}
%\dual \munit  & \defeq &  \bot  && \dual \bot & \defeq & \munit  
%\\
%\dual{(A \otimes B)} & \defeq & \dual A \parr \dual B && \dual{(A \parr B)} & \defeq & \dual A \otimes \dual B 
%\\
%\dual{(A \with B)} & \defeq & \dual A \oplus \dual B && \dual{(A \oplus B)} & \defeq & \dual A \with \dual B 
%\\
%\dual{(\oc {A})} & \defeq & \wn {\dual A} && \dual{(\wn {A})} & \defeq & \oc {\dual A}
%\\ 
%\dual{(\forall\tvar{X}.A)} & \defeq & \exists\tvar{X}.\dual A  && \dual{(\exists\tvar{X}.A)} & \defeq & \forall\tvar{X}.\dual A 
%\end{array} 
%\quad 
%\textit{and} \quad \dual{(\dual{\tvar{X}})} \defeq \tvar{X}
%\end{gather*}}%
%%
%
%
The multiplicative conjunction $A \otimes B$ (with unit $\munit$) is the type of a channel that 
communicates a name of type $A$ and a name of type $B$, offered by disconnected terms; 
it can be thought as an \qt{output.}  
%One interesting case is $A \otimes \lneg A$: it can be used to delay a 
%cut (a communication) between $A$ and $\lneg A$; this is evident since 
%the dual type $\lneg A \parr A$ can be implemented by $\inpb a x y \prls xy$, which 
%will \qt{connect} the two parts. 
The multiplicative disjunction $A \parr B$ (with unit $\bot$) is only different in that 
the communicated names can be offered by one term;  % (above we identified $x$ and $y$) %(\ie, they may have a dependency) 
this possibility of dependency makes it an \qt{input.}  
%The multiplicative units are $\munit$ and $\bot$, respectively, and they are assigned to nullary messages (signals). 
%
In a standard way, the additive conjunction $A \with B$ is an external choice (branching), 
and dually additive disjunction $A \oplus B$ is an internal choice (selection). 
%\ie, to 
%a name that can offer one of $A$ or $B$, but not both.   
%Dually, the additive disjunction $A \oplus B$ is the type of a name that will 
%communicate an internal choice (left: $A$, or right: $B$) to an external one.   
%%%%%%%%%%%%%%%%%We do not include additive units, since they are not crucial % from the computational perspective, 
%%%%%%%%%%%%%%%%%and the type $\topt$ causes complications in lazy reduction.\footnote{Because the context $\Ga$ 
%%%%%%%%%%%%%%%%%in $\vdash \Ga, \topt$ is inaccessible without commutations: it \qt{swallows} the outside.}
%
%The additive units are excluded for lack of useful function, and due to the complications 
%discussed in Remark~\ref{rema:additiveunits}. 
%
%Obviously, the binary connectives can be generalised to the $n$-ary case, but this is not 
%necessary for our study. 
%
%Their units are $\bot$ and $\tzero$ (which should not be confused with the inaction \nil). 
Ignoring modes, the exponential types $\oc A$ and $\wn A$  
can be understood as a decomposition of the \qt{shared} type in sessions:   
$\oc A$ %(\qt{of course} $A$) 
is assigned to a persistent term that offers $A$;  
%
%denotes the storage of a term 
%of type $A$ that can be used ad infinitum. 
%It is assigned to a persistent name, 
%and in fact it was Robin Milner's inspiration for the notation $\oc P$~\cite{milner:functions-as-processes}. 
dually, $\wn A$ %(\qt{why not} $A$) is the use of storage, or the invocation of persistent code, 
%and it 
can be assigned to any name with type $A$ 
so that it can communicate with %a term of type 
$\oc \lneg A$.  
%It can be understood as reading from memory, invoking a service, \etc 
%This observation hints at recursive types, which are discussed later. 
%
The second-order types $\forall\tvar X.A$ and $\exists\tvar X.A$ are standard,  
%The second-order type $\forall\tvar X.A$ (\qt{for all}) 
%is assigned to a name that inputs a type (for $\tvar X$) and a name used as 
%$A$ (allowing $\tvar X$ free in $A$). 
%%A term with this type is like universal abstraction, \ie, an input. 
%Dually, $\exists\tvar X.A$ (\qt{exists}) is the type of a name that outputs a type $B$ 
%and a name with type $A\subTP{B}{\tvar{X}}$. 
as is type substitution: 
$A\subTP{B}{\tvar{X}}$ stands 
for $A$ with $B$ for \tvar{X},
and $\dual{B}$ for $\dual{\tvar{X}}$.  
%We write $A\subTP{B}{\dual{\tvar{X}}}$ for 
%$A\subTP{\dual B}{\tvar{X}}$. 

\mypar{Judgements and interfaces}
A judgement $P \hast \Ga$ denotes that term $P$ can be assigned the 
{\em interface} $\Ga$. 
Interfaces, ranging over $\Ga,\De$, are sequences with possible repetition, defined by:
{\small \begin{gather*} 
\Ga  \quad ::= \quad   \emptyset \quad | \quad \Ga, a \tp   A 
\quad | \quad   \Ga, [a \tp A] \quad | \quad \Ga, \tvar X 
\end{gather*}}%
$a \tp  A$ is standard. 
A {\em discharged occurrence} $[a \tp A]$ indicates that $a$ has been used as $A$: 
%There are two possibilities, either a name is used by appearing in passive position, that is, 
%as the object of communication (or proof theoretically, as a premise to a connective), 
%or alternatively a name is used to effect a communication (in a cut between the conclusions of two connectives). 
%such occurrences are subject to scope restriction, and their presence 
it serves to protect linearity, since $a$ can no longer be used. 
$\Ga, \tvar X$ records that $\tvar X$ appears free in the term, 
ensuring freshness of type variables.  
$\vect a \tp \vect{A}$ stands for $a_1 \tp A_1, \ldots, a_n \tp  A_n$. 
%With $\vect a \tp \vect{A}$ we denote a sequence as above, but in that case the mode is always the same. 
$\wn \Ga$ stands for $\vect a \tp \vect{\wn A}$, \ie, $a_1 \tp \wn A_1, \ldots, a_n \tp  \wn A_n$. 
Similarly, $[\Ga]$ means $[\vect a \tp \vect{A}]$.  
%For any $\Ga\, (= \vect a \tp \vect A)$, we write $\lneg \Ga\, (= \vect a \tp \dual{\vect A})$ 
%for the orthogonal interface, \ie, for the same interface 
% with all types negated. 
Let \smm{\fn{a\tp A} = \fn{[a\tp A]} = a}   and \smm{\fn{\tvar X} = \emptyset}, 
plus the obvious definition for free type variables ($\ftv\Ga$).
%
%\mypar{Well-formedness} 
We consider \textit{well-formed interfaces}, 
in which  
%
%{\small \begin{mathpar}
%\dfrac{}{\wfg\emptyset} 
%\qquad
%\dfrac{\wfg \Ga \quad a \not\in \fn{\Ga}}   %\fn{\Ga}\cap\fn{\alpha \tp A}=\emptyset}
%     {\wfg \Ga, [a \tp A]}
%\qquad
%\dfrac{\wfg \Ga \quad a \not\in \fn{\Ga}}
%     {\wfg \Ga, a \tp  A}
%\qquad
%\dfrac{\wfg \Ga \quad \exists\sigma . \sigma(\Ga) = [\De], \Sigma \quad \tvar X \not\in\ftv \Sigma }{ \wfg \Ga, \tvar X }
%\qquad
%\dfrac{ \wfg \Ga \quad a \tp \wn A \in\Ga } %\Ga \!=\! \De, a \tp \emd A, \Theta \quad \mathsf{e} \in \set{\oc,\wn} }
%     { \wfg \Ga, a \tp \wn A }
%%\qquad \textit{In all type rules: } \:\:\framebox{$\dfrac{\cdots \quad \wfg{\Ga}}{P \hast \Ga}$}
%\end{mathpar}}%
%%
%%
%%
%The well-formedness conditions capture standard intuitions about linear occurrence.  
%Notably, 
only $a \tp \wn A$ %or %$a \tp \tmds \corect X A$ (but not both) 
can appear multiple times, but $a \tp \wni A$ cannot. % in $\Ga$.
%,  and similarly for $a \tp \mathfrak{c} \wn A$ 
%where $\mathfrak{c}$ is a critical modality.   
Moreover, ($\Ga, \tvar X$) is well-formed when $\Ga$ is well-formed 
and $\exists\sigma . \sigma(\Ga) = [\De], \Sigma$ such that $\tvar X \not\in\ftv \Sigma$.
For example in the \tsf{($\forall$)} rule the conclusion is $\Ga, [x \tp A], \tvar X, a \tp \forall \tvar X . A$ 
with $\tvar X$ possibly free in $A$. 
%Rules are omitted due to space restrictions. 

\mypar{Subtyping} 
The usual structural rules of \LL\ %, and even a form of dereliction, 
are incorporated into the relation $\Ga \subt \De$:%  
{\small \begin{gather*}
\Ga, [a\tp A] \subt \Ga, [a\tp \dual{A}] 
\qquad 
\Ga \subt \sigma(\Ga) 
%\qquad 
%\Ga, a \tp \bot \subt \Ga 
\qquad
\Ga, a\tp\wnm A \subt \Ga 
\\[4pt]
\Ga, a\tp\wn A \subt \Ga, a\tp\wn A, a\tp\wn A 
\qquad
\Ga, a \tp \wn A \subt \Ga, a \tp \wni A
\qquad
\Ga, a \tp \oci A \subt \Ga, a \tp \oc A
\end{gather*}}%
The first rule identifies the type of a discharged occurrence and its dual, 
matching a type annotation which may be  
$\newc{a \tp A}$ or $\newc{a \tp \dual A}$. 
%This simply reflects the symmetry of cut. 
Then we have {\em exchange}, %{\em bottom}, 
{\em weakening}, 
{\em contraction}. 
The last two axioms alter the mode: we can forget \mdi\ in $\wni A$, 
and dually we can record it on $\oc A$. 

\mypar{Typing rules} can be found in Fig.~\ref{fig:TSLL}. %
We type modulo structure equivalence, 
a possibility suggested by~\cite{milner:functions-as-processes} and used 
in~\cite{Caires:2010:SILL}. 
This is because associativity of \qt{$\,|\,$}
does not preserve typability, \ie, a cut between $P$ and $(Q \prls R)$ 
may be untypable as $(P \prls Q) \prls R$; % against $R$. 
\newc{a} causes similar problems.  

%In \tsf{(New\tvar X)} %, we have by well-formedness that $\tvar X \not\in\ftv{\Ga'}$ 
%%with $\sigma (\Ga) = [\De], \Ga'$, but 
%we require %the stronger condition 
%$\tvar X \not \in \ftv{\Ga}$, 
%%since we do not want $\tvar X$ to appear in the interface at all; this 
%which is stronger than well-formedness.  
%Whenever an interface contains $[a \tp A]$ or $\tvar X$  
%it will also contain at least one conclusion $c \tp C$, so we never derive the empty sequent,  
%as can be checked easily by inspection of the rules. 
%, but typing modulo 
%$\equiv$ resolves all these problems. 
%
%
\begin{figure}
{\small \begin{gather*}
\trule{New}{P \hast \Ga, [a\tp A] %[a\tp C] \qquad A \asymp C  %% the above introduces a free type variable...
                  }{ \newc{a\tp A} P \hast \Ga } 
\qquad
\trule{New \tvar X}{ P \hast \Ga, \tvar X \\ \tvar X \not\in\ftv{\Ga} }
                             { \newc{\tvar X}\, P \hast \Ga } 
\qquad 
\trule{Sub}{ \Ga \subt \De \\ P \hast \De }{ P \hast \Ga } 
\qquad 
\trule{Str}{ P \equiv Q \\ Q \hast \Ga }{ P \hast \Ga } 
\lns
\trule{Ax}{ \phantom{P} }{ ab \hast  a \tp A,  b \tp \dual A} 
\qquad 
\trule{Cut}{ P \hast \Ga, a \tp A \\ Q \hast \De, a \tp \lneg A 
                }{P \prls Q \hast [a \tp A], \Ga, \De } 
\qquad
\trule{OpenCut}{ P \hast \Ga, a \tp \ocm A \\ Q \hast \De, a \tp \wnm \lneg A }
                       {P \prls Q \hast  \Ga, \De, a \tp \ocm A }
\lns 
%\trule{Mix}{ P \hast \Ga \\ Q \hast \De }
%                       {P \prls Q \hast \Ga, \De}
\trule{CoMix}{ P \hast \Ga, \Theta \\ Q \hast \De,  \Theta  \\ \Theta \subseteq \set{ a \tp \wni A }
                 }
                       {P \prls Q \hast \Ga, \De, \Theta  
                       }
\qquad
\trule{\munit}{ \phantom{P} }{ \ovl a () \hast a\tp \munit}
\qquad  
\trule{\bot}{ P \hast \Ga}
          {a() \prls P \hast \Ga, a\tp \bot}  
\lns
\trule{\otimes}{ P \hast \Ga, b\tp   A \\ Q \hast \De, c\tp   B             
                       }{ \outs a {b,c} \prls P \prls Q 
                          \hast [b\tp A, c\tp B], \Ga, \De, a\tp A\otimes B }
\qquad
\trule{\parr}{P \hast \Ga, b\tp  A, c\tp  B}
                       { a(b,c) \prls P \hast [b\tp  A, c\tp  B], \Ga, a\tp A\parr B }
\lns
\trule{\oplus_1}{ P \hast \Ga, b\tp  A }
                        { \choiceI a 1 b \prls P \hast \denv{b\tp A}, \Ga, a\tp A \oplus B }
\qquad 
\trule{\with}{ P \hast \Ga, b\tp  A \\ Q \hast \Ga, c\tp  B \\ b, c \not\in\fn{\Ga}   }
                   {  \branch a { \methI 1 {b\tps A} P \orBra \methI 2 {c\tps B} Q } \hast \Ga, % \setminus \set{b,c}, 
                            a\tp A \with B  }
\lns
\trule{\forall}{ P \hast \Ga, b \tp A \quad \tvar X \not \in \ftv \Ga  
      						%% Not OK by well-formedness, not even [a : A{X}] should be here
                     }
                    { \inpb a {\tvar X} b \prls P \hast [b \tp A], \tvar X, \Ga, a \tp \forall\tvar X.A } 
\quad\:
\trule{TyAl}{ P\subTP{A}{\tvar X} \hast \Ga }{ \tyal X A \prls P \hast \Ga, \tvar X }
\quad\:  
\trule{\exists}{ P \hast \Ga, b \tp C \qquad C = A\subTP{B}{\tvar X} } 
                     { \outb a B b \prls P \hast [b \tp C], \Ga, a \tp \exists \tvar X . A }
\lns
%
%\trule{\oc}{ P \hast \wn \Ga, \set{x_i \tp A_i}_{i \in I} \\ \forall i \in I \,.\, x_i \not\in\fn{\wn\Ga} 
%                 %\\ \dom{\wn \Ga} \cap (\cup_{i \in I}{x_i}) = \emptyset 
%					\quad I \not = \emptyset, |I| > 1 \Rightarrow \md = \mdi 
%                }{ \oc a_i(x_i \tps A_i)_{i \in I}.P \hast \wn \Ga, %\setminus \set{x_i}_{i \in I}, 
%                     \set{a_i \tp \ocm A_i}_{i \in I} }
\trule{\oc}{ P \hast \wn \Ga, \set{x_i \tp A_i}_{i \in I} \\ \forall i \in I \,.\, x_i \not\in\fn{\wn\Ga} 
                 %\\ \dom{\wn \Ga} \cap (\cup_{i \in I}{x_i}) = \emptyset 
					\quad I \not = \emptyset \quad \forall i \geq 2 \, . \, \md_i = \mdi 
                }{ \oc a_i(x_i \tps A_i)_{i \in I}.P \hast \wn \Ga, %\setminus \set{x_i}_{i \in I}, 
                     \set{a_i \tp \ocmi i A_i}_{i \in I} }
\qquad 
\trule{\wn D}{ P \hast \Ga, b \tp A }
                    { \wnouts a b  \prl P  \hast [b \tp A], \Ga, a \tp \wnm A }
\end{gather*}}%
\caption{Linear Logic Typing with Multiparty Promotion}  
  \label{fig:TSLL}
\end{figure}
%
%
%
%The axiom and cut rules can be understood very easily. 
In \tsf{(Cut)} the name $a$ is discharged  
and can then be closed with \tsf{(New)}. 
%
%The \tsf{(Mix)} rule~\cite{Girard87} allows to compose independent (\ie, non-communicating) processes, 
%which is obviously very desirable for concurrent programming. 
%
$(\tsf{OpenCut})$ was added for two reasons.   
First, it is intuitive, since we are not 
required to close the name, \ie, to fix the number of clients of $\oc A$,    
departing from the de facto interpretation of \qt{cut as composition under name 
restriction.} %which was the de facto understanding until now. 
Second, it is needed for soundness. 
Take {\small $R \defeq \newc{a \tp \oc A} \blk{ ab \prl \wnouts a x \prl P \prl \oc a(y).Q }$} 
typed with \smm{R \hast \Ga, b \tp \oc A}. 
Using \tsf{(R-Ax)} we obtain \smm{R \R \wnouts b x \prl P\subs b a \prl \oc b(y).Q}, 
which is {\em only} typable with the same interface by using $(\tsf{OpenCut})$; % is used, 
with \tsf{(Cut)} we obtain $\Ga, [b \tp \oc A]$.\footnote{
Several works~\cite{PerezCPT12,deyoung_et_al:LIPIcs,Wadler12propsess,Caires:esop13} 
would not enjoy subject reduction if this example could be transferred: their cut rule 
requires $\newc b (\cdots)$, which is here missing. %, but  {\em sharing} of $a \tp \wn A$.
%works {\em only} in compositions under a bound name. 
%The culprit is {\em sharing} of multiple $\wn \lneg A$ against one $\oc A$, 
%in contrast to standard linear proofs (\PNS) in which \qt{contraction links} duplicate  
%terms of type $\oc A$ before each copy interacts. 
These works don't have \qt{\tsf{Mix}} (here: \tsf{(CoMix)}), which we used in the example; 
but this should be checked, since \qt{\tsf{Mix}} can be encoded with a new conclusion $c \tp \bot \otimes \bot$~\cite[p. 100]{Girard87}. 
%%but $\wn$-typed names  are shared % that do not reduce by copying the $\oc$-term (in contrast to \PNS), 
%%and axioms reduce similarly. % a similar reduction. 
%In~\cite{PerezCPT12,deyoung_et_al:LIPIcs,Caires:esop13}, 
%the problem is supposedly avoided by not allowing axioms to have an exponential 
%type, but then the interpretation is not accurate, and it is hard to see how 
%polymorphism works, since an axiom with types $\tvar X$, $\lneg \tvar X$ 
%can be instantiated with $\oc A$, $\wn \lneg A$.    
%In relation to~\cite{Wadler12propsess} it should be checked, 
%%but the geometry of \tsf{(Cut)}, \tsf{($\otimes$)} and \tsf{(Mix)} is very similar, 
%%and in fact 
%as it allows exponential axioms, 
}
%
%These rules are very simple, so we restrict ourselves to some comments. 
%
%First, deadlock, which is a form a cycle and therefore a manifestation of {\em inconsistency,} 
%is avoided because \tsf{($\otimes$)} splits the context into two {\em non-communicating} parts. 
%Which means that a cut between an output 
%\tsf{($\otimes$)} and an input \tsf{($\parr$)} will generate two independent cuts, \ie, it preserves acyclicity. 
%We create some intermediate axioms in reduction, but the essence remains.
%
%

Asynchronous messages can encode standard sessions (see~\cite{Demangeon:2011:FAS,deyoung_et_al:LIPIcs}): 
%
%Types $A \otimes B$ and its dual $\lneg A \parr \lneg B$ can be used 
%in the familiar sessions style, encoding continuations with one of the components 
%of output (the second in~\cite{Demangeon:2011:FAS,deyoung_et_al:LIPIcs}) and input. 
%
$\outb a b c$ with type $A \otimes B$ maps to  
the session type $\oc_{\tsf{s}} \lneg A.B$ {\em or} $\oc_{\tsf{s}} \lneg B.A$.  
Dually, $\inpb a x y$ with $\lneg A \parr \lneg B$ maps to $\wn_{\tsf{s}} \lneg A.\lneg B$ {\em or} $\wn_{\tsf{s}} \lneg B.\lneg A$.  
To write processes in standard sessions style, with reuse of names (\eg, $\ovl a b ; a(x); \nil$), 
we introduce abbreviations that use the second component for $a$'s continuation:%   
{\small \begin{gather*}
\begin{array}{rcl}
\ovl a b ; P & \defeq & \newc {x,y} \blk{ \outb a x y \prl xb \prl P \subs y a  }  
\\[2pt]
a \triangleleft l_k ; P  & \defeq & \newc x \blk{ \choiceI a k x \prl P \subs x a  }
\\[2pt]
\ovl a B ; P & \defeq & \newc {x} \blk{ \outb a {B} x \prl P \subs x a  }
\end{array}
\qquad
\begin{array}{rcl}
a(x) ; P & \defeq & \newc {x,y} \blk{  \inpb a x y \prl P \subs y a  } 
\\[2pt]
a \triangleright \set { l_i.P_i }_{i \in I} & \defeq & \branch a { \methI i {x_i} {P_i\subs {x_i}{a}} }_{i \in I} 
\\[2pt]
 a (\tvar X) ; P & \defeq & \newc {\tvar X,x} \blk{ \inpb a {\tvar X} x \prl P \subs x a  } 
\end{array}
\end{gather*}}%
It is easy to check that linear redices commute with all other redices, 
and therefore a \qt{real} prefix would not have any effect on computation except to make it more sequential.

The rule $(\oc)$ implements an extension of the logic: % with the rule:  
{\small 
\[
\frac{\wn \Ga, A}{\wn \Ga, \oc A} 
     \quad \textit{(promotion)}
\qquad \textbf{becomes} \qquad
\frac{\wn \Ga, A_1, \ldots , A_n}{\wn \Ga, \oc A_1, \ldots , \oc A_n} 
     \quad \textit{(multi-promotion)}
\]}%
% 
%which we can call \qt{multi-promotion.} 
Actually we need to employ some restrictions on this rule, which is 
why all conclusions except the first must have a $\mdi$-mode. 
Since there is no contraction for $\wni$,\footnote{More accurately: contraction of $\wni \lneg A_i$ is {\em multiplicative}.} 
all the $a_i \tp \wni \lneg A_i$ ($i \geq 2$) will come from terms with just one call 
to the session.\footnote{In the sense that two calls can never depend on each other.} 
The first conclusion, $a_1 \tp \ocmi 1 A_1$, can have standard mode ($\md_1 = \mde$), 
which allows a client's call with $a_1 \tp \wn \lneg A_1$ to be connected   
to (\ie, to depend on) other calls on $a_1$.  In this way we provide a 
{\em hook} for one client to participate in another instance of the same session, 
and this facilitates a form of {\em dynamic join}. 
We return to this concept in the first example. 

Finally, the sidecondition in $(\with, \oc)$ % / $(\oc)$ 
forbids premises from having multiple copies of a name (\eg, $x \tp \wn A, x \tp \wn A$) 
which should be removed in the conclusion; % are treated correctly. %, % in the conclusion, 
%which is possible with , it will be \qt{contracted} (by $\subt$) 
%so as not to appear in the conclusion. 
%
%avoids situations such as 
%%For example, consider $(\with)$ with 
%having a left premise 
%$P \hast \Ga', c \tp B, b \tp \wn A',b \tp \wn A'$, obtaining $\Ga', c \tp B, b \tp \wn A'$. % in the conclusion. 
essentially it forces contractions (by $\subt$). %, which are always possible. %, and the problem disappears. 
Other rules are immune by  the well-formedness of 
$\Ga, [a \tp A]$\footnote{It is subtle but due to the variable convention, $(\with)$ is actually immune too; the condition serves for clarity.}. 
%

%
%cut against a term of type $\wn \lneg A_1, \ldots , \wn \lneg A_n$ can form a cycle. 
%To see why, notice that the $A_i$ can have dependencies, and similarly for the $\lneg A_i$, \ie, a communication 
%would correspond to a use of multi-cut, which is known to allow cycles. 
%This is the purpose of the $\mdi$-mode, for all but one of the conclusions: it forces each $\wni \lneg A_i$ 
%to come from an {\em independent} term, so the synchronised communication can in fact be 
%decomposed into standard cuts, and no cycle can be formed. 
%In particular, contraction of $\wni \lneg A_i$ is {\em multiplicative}. 
%The first conclusion, $\oc A_1$, can have standard mode, which allows the $\wn \lneg A_1$ to be connected 
%to other instances of the same name, \ie, we provide a hook for one client to participate in another session 
%with the same server, and this facilitates a form of dynamic join. %Details are omitted. % due to lack of space. 
%%For similar reasons we need to restrict contraction of \wni-types, 
%%in particular it is only allowed in \tsf{(CoMix)}: two processes can share 
%%up to one name of type $a \tp \wni A$.
%%
%%The combination of sharing (multiple $a \tp \wn A$) with synchronisation 
%%makes the system non-deterministic,  
%%%
%%%Most logicians would object to such a rule, at least with our interpretation of it,  
%%%since confluence is really lost. 
%%%However, other properties of interest are preserved, and one can legitimately ask 
%%but can we demand confluence in concurrent programming? 

%

\mypar{Expressiveness \& Properties}\label{sec:prop} % 
%
%
%\textbf{DROP units !!}
%
The system is an extension of \PNS,
%\footnote{If we restrict replication to have a singleton input, we obtain (almost) standard \PNS.}  
in process form, 
so it can encode System F, inductive sessions (using second-order features), etc. 
%, inductive data, \etc, in a standard way. 
%Inductive sessions can be encoded using second-order features.
Due to space limitations we only show two examples:  
(a) how shared channels can be simulated with synchronisation; 
(b) the (two Buyer, one Seller) protocol from~\cite{HondaK:mulast}. 
%More details will be reported in \url{http://www.di.fc.ul.pt/~dimitris/PN/}.

\mypar{a) channels} Non-determinism can be  
expressed by sharing a channel between multiple competing processes 
trying to send and receive messages. This is impossible with existing logical sessions 
systems, and more generally if we follow the logic 
%, since it implies the loss of confluence and, more generally, cannot 
%be typed if we follow the logic 
\qt{by the book.} %\footnote{If we restrict replication to have a singleton input, we obtain (almost) standard \PNS.}  
A channel $a$ with \tsf{i/o} type $(A,\lneg A)$, \ie, that exports two complementary capabilities $A$ and 
$\lneg A$, can be encoded 
by the two names $a_1$ %(input) 
and $a_2$ %(output) 
in 
$\oc a_{1}(x\tps A) \, a_{2}(y\tps \lneg A). xy \hast a_1 \tp \oc A, a_2 \tp \oci \lneg A$. 
%
%:%
%{\small \begin{gather*}
%%
%\tsf{Ch}[a \tp \tsf{i}(A)] \defeq \oc a_{1}(x\tps A) \, a_{2}(y\tps \lneg A). xy 
%%
%\end{gather*}}%
%
%The channel exports the interface $a_1 \tp \oc A, a_2 \tp \oc \lneg A$. 
The channel is used by terms with $a_1 \tp \wn \lneg A$ or $a_2 \tp \wni A$, and there can be multiple 
instances of each, giving rise to critical (non-deterministic) pairs. % and inducing non-determinism. 
Moreover, $A$ can be linear, \ie, we can communicate linear values through shared channels, 
which is a novel feature. 
%In fact, the above types can be abstracted, and we can instantiate channels 
%of type $\forall \tvar X . \oc \tvar X \parr \oc \lneg \tvar X$. 
For example: 
{\small %
\begin{gather*}
\wnouts {a_1} {b_1} \prl \wnouts {a_2} {b_2} \prl \wnouts {a_2} {b_3} \prl \oc a_{1}(x\tps A) \, a_{2}(y\tps \lneg A). xy \prl P \prl R \prl S 
\\ 
\R^{(a)} 
b_1b_2 \prl  \wnouts {a_2} {b_3} \prl \oc a_{1}(x\tps A) \, a_{2}(y\tps \lneg A). xy \prl P \prl R \prl S 
\\
\R^{(b)} b_1b_3 \prl  \wnouts {a_2} {b_2} \prl \oc a_{1}(x\tps A) \, a_{2}(y\tps \lneg A). xy \prl P \prl R \prl S 
\end{gather*}}%
First, note that confluence is lost: assume $P, R, S$ cannot reduce and it becomes obvious. 
The graphical notation with a reduction of the first possibility is depicted below.  

\noindent {\small 
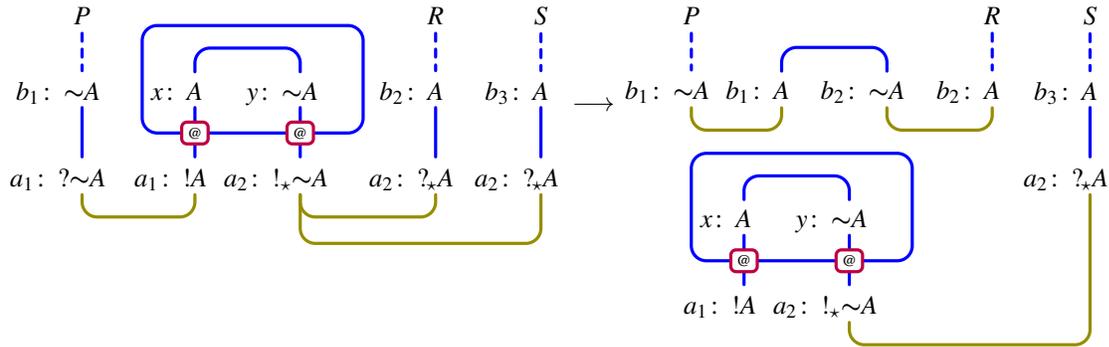
\begin{tikzpicture}%

\coordinate (ca1) at (1,1);
\coordinate (ca2) at (2.5,1);
\coordinate (ca3) at (3.9,1);

\coordinate (ca21) at (5.7,1);
\coordinate (ca22) at (7.1,1);

% dereliction 
%
\UnaLink{a_1}{b_1}{\wn \lneg A}{\lneg A}{ca1}{a1}  ;
\UnaLink{a_2}{b_2}{\wni A}{A}{ca21}{a21}  ;
\UnaLink{a_2}{b_3}{\wni A}{A}{ca22}{a22}  ;

\path (a1-T) +(0,1cm) coordinate (nP);
\Snode{P}{nP}{P} 
\draw[premise] (P.south) -- (a1-T) ; 

\path (a21-T) +(0,1cm) coordinate (nR);
\Snode{R}{nR}{R} 
\draw[premise] (R.south) -- (a21-T) ; 

\path (a22-T) +(0,1cm) coordinate (nS);
\Snode{S}{nS}{S} 
\draw[premise] (S.south) -- (a22-T) ; 

% Promotion 
%
%
\OkayLink{a_1}{x}{\oc A}{A}{ca2}{a2}
\OkayLink{a_2}{y}{\oci \lneg A}{\lneg A}{ca3}{a3}
\AxLinkB{a2-T}{a3-T}{0cm}{}

%\path (a2-T) +(0,1cm) coordinate (nQ);
%\Snode{Q}{nQ}{Q}     
%\draw[premise] (Q.south) -- (a2-T) ; 
%
%\path (nQ) +(0.3cm,0) coordinate (nT);
%\RecVarLink[1.0cm]{nT}{T}{\mathbf{t}}{0}{} 
%
\coordinate (aboveAx) at ($(Ax-a2-T-a3-T-M) + (0,0.3cm)$) ; 
%\coordinate (rightT) at ($(nT.east) + (0.1cm,0)$) ;
%
%
\aFittedBox{theBox}{(a2-T) (a3-T) (aboveAx) (a2-at-crd) (a3-at-crd)}{link, inner sep=0pt}

\CutLink{a1}{a2}{0pt}{}
\CutLink{a3}{a21}{0pt}{}
\CutLink{a3}{a22}{10pt}{}

%
%  now the reduct 
% 

\coordinate (redu) at (7.8,2);
\Snode{\bfs\R}{redu}{arr} 

\coordinate (nb1) at ($ (a2-T) + (6.6cm,0cm) $) ;
% 1: name, 2: Type, 3: origin, 4: node name
\Tnode{b_1}{\lneg A}{nb1}{newb1}

\coordinate (nb11) at ($ (nb1) + (1.2cm,0) $) ;
\coordinate (nb12) at ($ (nb11) + (1.4cm,0) $) ;
\Tnode{b_1}{A}{nb11}{newb11}
\Tnode{b_2}{\lneg A}{nb12}{newb12}
\AxLinkB{newb11}{newb12}{0cm}{}

\coordinate (nb21) at ($ (a21-T) + (7.4cm,0cm) $) ;
% 1: name, 2: Type, 3: origin, 4: node name
\Tnode{b_2}{A}{nb21}{newb21}

\path (nb1) +(0,1cm) coordinate (nP1);
\Snode{P}{nP1}{P1} 
\draw[premise] (P1.south) -- (newb1) ; 

\path (nb21) +(0,1cm) coordinate (nR1);
\Snode{R}{nR1}{R1} 
\draw[premise] (R1.south) -- (newb21) ; 

\CutLink{newb1}{newb11}{0pt}{}
\CutLink{newb12}{newb21}{0pt}{}

\coordinate (newca22) at ($ (ca22) + (7.3cm,0cm) $) ; 
\UnaLink{a_2}{b_3}{\wni A}{A}{newca22}{newa22}  

\path (newa22-T) +(0,1cm) coordinate (newcS);
\Snode{S}{newcS}{nS} 
\draw[premise] (nS.south) -- (newa22-T) ; 

% Promotion 
%
%
\coordinate (newca2) at ($ (ca2) + (7.3cm,-1.7cm) $) ; 
\coordinate (newca3) at ($ (ca3) + (7.3cm,-1.7cm) $) ;

\OkayLink{a_1}{x}{\oc A}{A}{newca2}{newa2}
\OkayLink{a_2}{y}{\oci \lneg A}{\lneg A}{newca3}{newa3}
\AxLinkB{newa2-T}{newa3-T}{0cm}{}

\coordinate (aboveNewAx) at ($(Ax-newa2-T-newa3-T-M) + (0,0.3cm)$) ; 
%\coordinate (rightT) at ($(nT.east) + (0.1cm,0)$) ;
%
%
\aFittedBox{theBox}{(newa2-T) (newa3-T) (aboveNewAx) (newa2-at-crd) (newa3-at-crd)}{link, inner sep=0pt}
\CutLink{newa3}{newa22}{0pt}{}

%\coordinate (nx1) at ($ (a21-T) + (8cm,0cm) $) ;

%\AxLinkB{a2-T}{a3-T}{0.2cm}{}

\end{tikzpicture}%
}%
%
%\mypar{Translating \pic} is not direct because our system does not admit processes that deadlock or diverge. 
%To get a close correspondence (up to a reasonable weak equivalence) we need $\otimes = \parr$, and to 
%remove the context split in \qt{Cut} rules, 
%but this would be a logical atrocity. 
%

\NI It is possible that $P$ has another call to $a_1$, but by the restriction on 
$\wni$-types there cannot 
be a \qt{trip} from $b_2$ to $a_2$, as this would lead to a cycle.  
Concretely, if $R$ has another call to $a_2$, then it is from a part disconnected to $b_2$, 
and similarly for $S$; see \tsf{(CoMix)}.%
\footnote{In general, derelictions can be connected through their premises; try with two copies of $a\tp \wn( A \oplus \lneg A)$.}  
%Without the special conditions on $\mdi$-types, we admit cy 
%For example, If there is another call to $a$ (\tsf{CoMix}) 

\mypar{b) multiparty interactions} 
The (two Buyer, one Seller) protocol from~\cite{HondaK:mulast} 
is shown below, with insignificant adaptations, using the previously explained abbreviations (we omit some signals for $\munit/\bot$): 
{\small \begin{gather*}
\begin{array}{rcl}
\tsf{Buyer1} & \defeq &  \newc{b_1} \left( \wnouts {a_1} {b_1} \prl 
          \ovl{b_1} {\,\text{\qt{The Art of War}}} ; b_1 (quote) ; \newc z \blk{ \ovl{b_1} z; \nil \prl\ovl{z} \, quote / 2 ; P_1  } \right)
\\[2pt]
\tsf{Buyer2} & \defeq &  \newc{b_2} \left(  \wnouts {a_2} {b_2} \prl 
   b_2(quote); b_2(z); z(contrib); b_2 \triangleleft \tsf{ok};  \ovl{b_2} \,\text{\qt{SW12 3AZ}} ; b_2(date); P_2 \right)
\\[2pt]
\tsf{Seller} & \defeq &
    \oc a_1(x_1) \, a_2(x_2) . \left( 
      					\begin{array}{l}
      					x_1(title) ; \ovl{x_1} \, \text{\EUR{20}} ; \ovl{x_2} \, \text{\EUR{20}} ;
                            x_1(z); \ovl{x_2} z; 
                            \\[2pt]
                            \qquad x_2 \triangleright \set{ \tsf{ok}.x_2(address); 
                              \ovl{x_2} \, \text{\qt{7/Feb}}; Q  \orBra \tsf{quit}.\nil } 
      					\end{array}
                         \right) 
\end{array}
\end{gather*}}
\NI We note that the simplicity 
of the example has not been sacrificed, compared to the code in~\cite{HondaK:mulast}. 
One difference is that we passed $z$ from \tsf{Buyer1} to \tsf{Buyer2} {\em through} \tsf{Seller} 
using $ x_1(z); \ovl{x_2} z$, when in~\cite{HondaK:mulast} all names are known to all participants. 
We do not employ the {\em global types} of~\cite{HondaK:mulast}, 
but there is a {\em proof net} for $\tsf{Buyer1} \prl \tsf{Buyer2} \prl \tsf{Seller}$,
 not shown due to space constraints, and we postulate that:%
\begin{center}
 \framebox{{\em The proof net can serve as an alternative notion of global type.}}
 \end{center} 
% 
%The lack of {\em global types}~\cite{HondaK:mulast} is perhaps a disadvantage, 
%Finally, our system enjoys stronger properties, such as termination. 
% and is also polymorphic,  
%so it can serve as a strong 
%foundation for logical multiparty sessions. 
%This is reinforced by the fact that global types help obtain stronger properties compared 
%to session types, but these ar
%Multiparty synchronisation seems related to the Join calculus~\cite{DBLP:conf/popl/FournetG96}, and to our knowledge 
%this relationship has not yet been investigated; we leave it as future work. 

\mypar{Outline of results} 
The expected soundness result for reduction, 
$P \hast \Ga$ and $P \R P'$ implies $P' \hast \Ga$, 
 is obtained in a standard way, but fails without the \mdi-mode. 
Strong Normalisation (\textbf{sN}), \ie, $P \hast \Ga$ implies that \textit{all} 
reduction sequences from $P$ are finite, is shown by an adaptation of the 
reducibility candidates technique from~\cite{Girard87}. 
The loss of confluence complicates the proof, which is in fact obtained 
for an extended (confluent) reduction relation using a technique of~\cite{Ehrhard2010606}, 
from which we derive as a corollary the result. 
For \sN\ we prove the (initially) stronger property of {\em reducibility}~\cite{Girard87}, 
which can also serve as a very strong progress guarantee. 

A Curry-Howard correspondence can be obtained easily for a fragment of the language. 
For the multiplicative, additive, and second-order cut-elimination we only 
need to perform extra axiom cuts (\ie, substitutions). 
For exponentials, we restrict replications to a single input % (\ie, without synchronisation), 
and simulate the actual copying (with contraction links) that takes place 
in \PNS\ with sharing and sequentialised cut-elimination steps. 
Indeed, there is still a loss of parallelism 
compared to standard \PNS, but the term language is more realistic. % and clearly remains polynomialy equivalent.  
We show just one case of cut-elimination, the cut  ($\otimes$ --- $\parr$), implemented by 
$\outs a {b,c} \prl \inps a {x, y} \R bx \prl cy$, adding appropriate contexts ($P, Q, R$): 
%has the following proof net presentation 
%(we add appropriate contexts): 

 {\small 
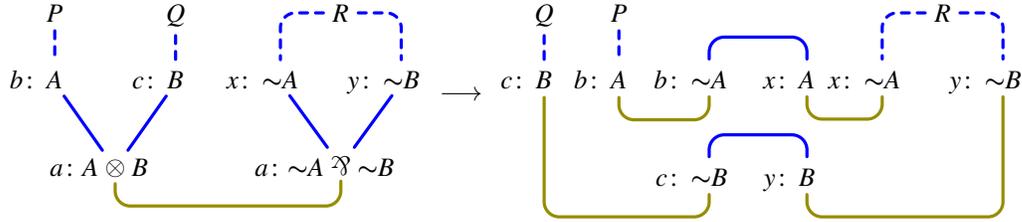
\begin{tikzpicture}%

\coordinate (ca1) at (1,1);
\coordinate (ca2) at (4,1);

%
% par link 
%
\BinLink{a}{\lneg A}{\parr}{\lneg B}{ca2}{a2} \Tnode{x}{\lneg A}{a2-L}{x} \Tnode{y}{\lneg B}{a2-R}{y}
\path (a2) +(0,2cm) coordinate (nR) ;
\Snode{R}{nR}{R} 
\draw[premise] (R.west) -| (x) ; 
\draw[premise] (R.east) -| (y) ;

%
% tensor link  
%
\BinLink{a}{A}{\otimes}{B}{ca1}{a1}  \Tnode{b}{A}{a1-L}{b} \Tnode{c}{B}{a1-R}{c}
\pgfextractx{\userDimA}{\pgfpointanchor{a1-L}{center}} 
\pgfextractx{\userDimB}{\pgfpointanchor{a1-R}{center}}
\pgfextracty{\userDimC}{\pgfpointanchor{nR}{center}}
\coordinate (nP) at ($(\userDimA,\userDimC)$);
\coordinate (nQ) at ($(\userDimB,\userDimC)$);
\Snode{P}{nP}{P} 
\Snode{Q}{nQ}{Q} 
\draw[premise] (P.south) -- (b) ; 
\draw[premise] (Q.south) -- (c) ;

\CutLink{a1}{a2}{0pt}{}

%
%  now the reduct 
% 

\coordinate (redu) at (5.6,2);
\Snode{\bfs\R}{redu}{arr} 

% R 
%
\path (a2) +(8cm,2cm) coordinate (nR1) ;
\path (x) +(8cm,0cm) coordinate (cx1) ;
\path (y) +(8cm,0cm) coordinate (cy1) ;
\Tnode{x}{\lneg A}{cx1}{x1}  \Tnode{y}{\lneg B}{cy1}{y1}
\Snode{R}{nR1}{R1} 
\draw[premise] (R1.west) -| (x1) ; 
\draw[premise] (R1.east) -| (y1) ;

% P, Q  
%
\path (nP) +(7.5cm,0cm) coordinate (nP1) ;
\path (nQ) +(4.9cm,0cm) coordinate (nQ1) ;
\path (a1-L) +(7.5cm,0cm) coordinate (nb1) ;
\path (a1-R) +(4.9cm,0cm) coordinate (nc1) ;

\Snode{P}{nP1}{P1} 
\Snode{Q}{nQ1}{Q1} 
\Tnode{b}{A}{nb1}{b1} \Tnode{c}{B}{nc1}{c1}
\draw[premise] (P1.south) -- (b1) ; 
\draw[premise] (Q1.south) -- (c1) ;

% Axioms 
%
% b-x
\path (cx1) +(-1cm,-0cm) coordinate (nax1-R) ;
\path (cx1) +(-2.3cm,-0cm) coordinate (nax1-L) ;
\AxLink{b}{\lneg A}{x}{A}{nax1-L}{nax1-R}{0.0cm}{} 
% c -y 
\path (cx1) +(-1cm,-1.3cm) coordinate (nay1-R) ;
\path (cx1) +(-2.3cm,-1.3cm) coordinate (nay1-L) ;
\AxLink{c}{\lneg B}{y}{B}{nay1-L}{nay1-R}{0.0cm}{}
%
% Cuts
%
\CutLink{b1}{Ax-nax1-L-nax1-R-L}{0pt}{}
\CutLink{Ax-nax1-L-nax1-R-R}{x1}{0pt}{}
\CutLink{c1}{Ax-nay1-L-nay1-R-L}{0pt}{crossing cut}
\CutLink{Ax-nay1-L-nay1-R-R}{y1}{0pt}{}

\end{tikzpicture}}

%IDEAS: 
%
%1) Non-deterministic selection (exists in logic)
%
%2) Dependent types for accepting an arbitrary finite number of requests. 
%This requires work since the "n" clients need to be handled uniformly; 
%the existing inductive abilities of proof nets may be enough. 
%
% each !A is actually !A[n] or !A[n < m] for some limit
%
% We then need a "function"  A[n] --o B to handle it. 
% This can be understoof as A x .... x A --o B   i.e. with n elements in tensor.... but cannot predict... list better.
%
% Maybe allow to iterate and "accept" more and more as required, maybe controlled by session. 
% Eventually need to release replication... in fact uniformity is at stake...
%
% Maybe introduce (full) COMPREHENSION which is possible in the logical system 
% This way we can treat sets of clients generically.
% See Girard's Book
%

%\input{PN-CUT-ELIM}

\section{Conclusion}\label{sec:related}

We claim that our language is simpler and proof-theoretically 
more appealing than related works such as~\cite{Caires:2010:SILL}: 
structured interactions take place as expected ({\em fidelity}), 
but parallelism is not inhibited by the use of prefix, which cannot anyway alter the result 
in a deterministic 
setting. % (\ie, modulo our synchronisation ability). 
It is really a question of \PNS\ {vs.} sequent proofs, and in logic 
the first are almost always preferable. 
%and both are confluent using lazy reduction. 
Even with synchronisation and the induced non-determinism, 
the system we propose retains good properties,  
%of the logical interpretation except for confluence, 
%and to our knowledge this 
for example it seems to be the first 
approach to multiparty behaviours that enjoys strong normalisation. 
%Related works that follow~\cite{Caires:2010:SILL} can only express binary sessions. 
Finally, our notion of {\em proof net as global type} seems to be a reasonable solution for 
logically founded multiparty sessions.

In relation to Abramsky's interpretation~\cite{Abramsky93CILL},  
it is close to proof nets {\em with boxes}, \ie, to a completely synchronous calculus. 
Moreover, it is not so friendly syntactically, it does not have a notion of bound name, 
copying of exponentials is explicit (no {\em sharing}), and of course it is completely deterministic. 
%
%(In a \pic\ setting, \sN\ has been obtained using a variation of reducibility candidates in~\cite{Yoshida:2004:SN}, 
%but again this system is confluent.)
% and hence also 
%progress: if a session cannot proceed, it is because it requires to be linked to a larger context.  
%
%Our synchronisation mechanism seems somewhat related to Join patterns~\cite{DBLP:conf/popl/FournetG96}, 
%but for now we leave this investigation as future work. 
%
An interesting future direction would be to obtain a {\em light} variation of 
our system, \eg, following~\cite{Girard95lightlinear}. 
Then we could speak of implicit complexity for multiparty sessions, similarly to 
what has been done in~\cite{DBLP:journals/corr/abs-1108-4467} for binary sessions. 
Due to space restrictions, more examples and all proofs have been omitted. 
These will appear in a longer version, see \url{http://www.di.fc.ul.pt/~dimitris/}.

%\nocite{*}
\bibliographystyle{eptcs}
\bibliography{sessions}
\end{document}